\documentclass[aps,jcp,superscriptaddress,amsmath,amssymb]{revtex4-2}

\usepackage{comment}
\usepackage{tikz}
\usetikzlibrary{trees}
\usetikzlibrary{decorations.pathmorphing}
\usetikzlibrary{decorations.markings}
\usepackage{tikz-cd}
\usetikzlibrary{arrows,arrows.meta}
\usetikzlibrary{decorations.pathreplacing,snakes}
\tikzset{
    photon/.style={decorate, decoration={snake,segment length=1.5mm}, draw=black},
    coulomb/.style={dotted},
    electron/.style={draw=black, postaction={decorate},
        decoration={markings,mark=at position .55 with {\arrow[draw=black]{>}}}}, 
    gluon/.style={decorate, draw=magenta,
        decoration={coil,amplitude=4pt, segment length=5pt}},
    boundelectron/.style={thick, double},
    transverse/.style={dashed},
    marrow/.style={decoration={markings,mark=at position 0.5 with {\arrow{#1}}}, postaction=decorate}    
}
\usepackage{graphicx}
\usepackage{dcolumn}
\usepackage{bm}
\usepackage{rotating}

\newcolumntype{.}{D{.}{.}{8}}

\usepackage[utf8]{inputenc}
\usepackage{physics}
\usepackage{amsmath, amssymb}
\usepackage{tabularx,booktabs,array,dcolumn,multirow}
\usepackage{setspace}
\usepackage{vmargin}
\usepackage{xcolor}
\usepackage{graphicx,psfrag,subfigure}

\usepackage{float}
\usepackage[all]{xy}

\usepackage{bm}
\usepackage{mathtools}
\usepackage{physics}
\usepackage[english]{babel}

\newcommand{\bos}[1]{\boldsymbol{#1}}
\newcommand{\mr}[1]{\mathrm{#1}}

\def\Eh{E_\mr{h}}

\def\iim{\mr{i}}
\def\eem{\mr{e}}

\def\nopair{\text{np}}

\def\DC{\text{DC}}
\def\DCB{\text{DCB}}

\def\ntwo{^{[2]}}
\def\nthree{^{[3]}}

\def\four{^{[4]}} 
\def\sixteen{^{[16]}} 

\def\bA{\vb*{A}}
\def\bv{\vb*{v}}
\def\bp{\vb*{p}}
\def\br{\vb*{r}}

\def\bs{\vb*{s}}

\def\balpha{\vb*{\alpha}}
\def\bsigma{\vb*{\sigma}}

\def\bnabla{\boldsymbol{\nabla}}

\def\buA{\underline{\vb*{A}}}

\def\si{\sigma}

\def\epsi{\varepsilon}

\def\pupu{\scalebox{0.65}{$++$}}
\def\pumi{\scalebox{0.65}{$+-$}}

\def\mimi{\scalebox{0.65}{$--$}}

\def\pp{{++}}

\def\som{Supplementary Information}

\def\tC{\text{C}}
\def\tB{\text{B}}

\def\tCB{\text{CB}}
\def\tT{\text{T}}

\def\inst{\text{I}}
\def\epsi{\varepsilon}
\def\Acal{{\cal{A}}}
\def\Pcal{{\cal{P}}}
\def\Vcal{{\cal{V}}}
\def\VQED{{\cal{V}}_\mr{QED}}
\def\Kcal{{{K}}}

\def\nulla{{(0)}}

\def\Vcale{\mathcal{V}_\epsilon}
\def\Kcal{\mathcal{K}}

\definecolor{carnelianred}{cmyk}{0, 0.864, 0.837, 0.274}
\definecolor{harvestgold}{cmyk}{0, 0.330, 0.981, 0.133}
\definecolor{patricksblue}{cmyk}{0.826, 0.644, 0, 0.525}
\definecolor{arany}{cmyk}{0, 0.295, 0.988, 0.003}
\definecolor{red}{cmyk}{0, 0.785, 0.785, 0.176}
\definecolor{blue}{cmyk}{0.680, 0.551, 0, 0.423}

\definecolor{gray}{rgb}{0.5,0.5,0.5}

\usepackage[unicode]{hyperref}
\usepackage{soul}
\hypersetup{
   unicode=true,          
   plainpages=false,
   colorlinks=true,       
   linkcolor=black,          
   linkcolor=blue,          
   citecolor=blue,        
   urlcolor=blue           
}

\usepackage{url}

\begin{document}

\title{%
Bound-state relativistic quantum electrodynamics: a perspective for precision physics with atoms and molecules
}

\author{\'Ad\'am Nonn}
\author{\'Ad\'am Marg\'ocsy}
\author{Edit M\'atyus} 
\email{edit.matyus@ttk.elte.hu}
\affiliation{ELTE, Eötvös Loránd University, Institute of Chemistry, 
Pázmány Péter sétány 1/A, Budapest, H-1117, Hungary}

\date{\today}

\begin{abstract}
\noindent %
Precision physics aims to use atoms and molecules to test and develop the fundamental theory of matter, possibly beyond the Standard Model. Most of the atomic and molecular phenomena are described by the QED (quantum electrodynamics) sector of the Standard Model. 
Do we have the computational tools,  algorithms, and practical equations for the most possible complete computation of atoms and molecules within the QED sector? What is the fundamental equation to start with? Is it still Schrödinger's wave equation for molecular matter, or is there anything beyond that?
This paper provides a concise overview of the relativistic QED framework and recent numerical developments targeting precision physics and spectroscopy applications with common features with the robust and successful relativistic quantum chemistry methodology.
\end{abstract}

\maketitle

\section{Introduction \label{ch:intro}}
In the Schrödinger equation ($H\psi=E\psi$) (1926), the Hamiltonian is the sum of the kinetic and potential energy ($H=T+U$) \cite{Sc26}. From special relativity (1905) it is known, that the energy of a moving mass $m$ with momentum $\bp$ is given by the Einstein energy relation: $E^2=\bp^2c^2+m^2c^4$ (where $\bp$ is the relativistic three-momentum $\bp=\gamma(\bv) m\bv$). This relativistic energy is the sum of the rest energy and the generalized kinetic energy, which reduces to the classical kinetic energy ($\frac{1}{2}m\bv^2$) for velocities much lower than the speed of light in vacuum ($c$) \cite{Ei05}. As the total energy can also include the potential energy due to an external potential, the relativistic Hamiltonian, in general, is $H=c\sqrt{\bp^2+m^2c^2}+U$. Paul Dirac wanted to rewrite `the square root' such that the Hamiltonian is linear in momentum. When he succeeded, the result was the Dirac equation (1928) \cite{Di28a,Di28b}:
\begin{align}
    h(\bp) \phi
    =
    \qty(c\balpha\bp + \beta mc^2 + U) \phi
    =
    E \phi \ ,
\end{align}
where $h(\bp)$ is the one-particle Dirac Hamiltonian, $\beta$ and $\balpha$ together are the Dirac matrices, which are complex valued 4$\times$4 matrices, and $\phi$ is the 4-component Dirac spinor, which represents massive, spin-1/2 fermions.

The Dirac equation is the relativistic description of such spin-1/2 fermions. It can be solved for the one-electron atom, where $U$ is the time-independent external electric potential energy generated by the infinite-mass nucleus \cite{Go28}. Dirac also named the relativistic treatment of electrons Quantum Electrodynamics (QED) \cite{Di27}.

The existence of negative-energy solutions to the Dirac equation predicted the existence of the positron, the antiparticle of the electron \cite{Di28a}. The original interpretation was that in vacuum, all negative-energy states are occupied by undetectable electrons. Then, if an extra electron was added, it would occupy a positive-energy state and be detectable as an electron, while if an electron was removed from a negative-energy state, the resulting `hole' could be detectable as a positron (1930) \cite{Di30}. Richard Feynman correctly reinterpreted the negative-energy solutions of the Dirac equation by using a Green's function formalism, and by introducing the Feynman propagator (1949). In a Green's function formalism, finding the Green's function of an operator is equivalent to solving its corresponding eigenvalue equation. The Feynman propagator is one of the possible choices for the Green's function corresponding to the Dirac equation, one in which positrons are mathematically treated as negative-energy electrons `travelling' backwards in time (see also Eqs.~\eqref{eq:FSpropag}--\eqref{eq:FSpropag2}) \cite{Fe49b,Fe49a}. This is the correct interpretation, and this treatment of positrons was a crucial milestone in the development of QED.

Today, the best (most accurate) physical theory describing reality on the smallest scale is the Standard Model, which is described in the language of Quantum Field Theories (QFTs). The sector within it describing electromagnetism is QED.

In QED, one of the basic cases studied is the scattering of two free particles. A scattering event has a (purposefully specific) definition: in a vacuum, two free particles approach the `interaction window' from the infinitely distant past ($t_\text{init}=-\infty$); they only interact over a very short distance and a very short duration of time (they only interact within the `interaction window'); then they leave the interaction, and remain free particles travelling into the infinitely distant future ($t_\text{fin}=\infty$). We are always allowed to take this limit of infinitely distant past and future, as the interaction window is very small. This definition makes the treatment relatively simple. An advantageous mathematical transformation can be performed, after which the calculation of the scattering amplitude (the probability amplitude that the scattering event will take place) only requires the solutions of the free Dirac equation: $\qty(c\balpha\bp + \beta mc^2)\phi=E\phi$. This transformation is called the interaction picture (or Dirac picture) \cite{Di27}.

\begin{figure}
  \centering
  \includegraphics[scale=0.85]{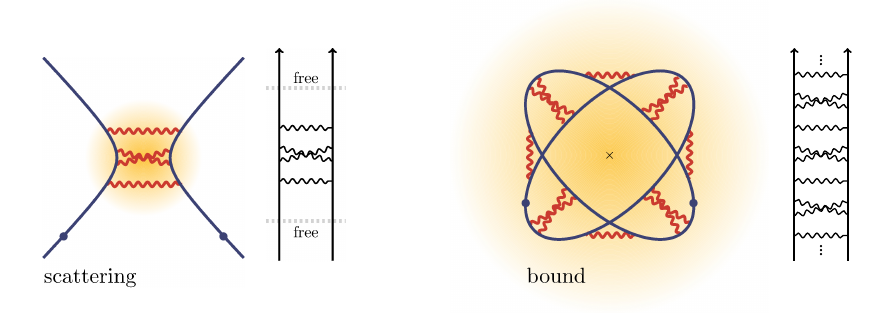}
  \caption{%
    Bound-state QED is challenging: the bound particles interact through an infinite series of interaction events during the infinite lifetime of the state. At the same time, scattering events are limited, in space and time, to a few interactions.
    \label{fig:bQED}
  }
\end{figure}

In this treatment, the calculation of the scattering amplitude involves calculating the matrix element of the $S$ matrix $\langle f|S|i\rangle$. $S$ is a time evolution operator, which evolves the initial state $i$ from the infinitely distant past to the infinitely distant future.
The matrix element $\langle f|S|i\rangle$ therefore gives the probability amplitude that a given initial state $i$ evolves into a given final state $f$. This matrix element is written as (a finite-order approximation to) an infinite series \cite{Dy49a,Dy49b,Dy52}.
Each term in this series can be represented by Feynman diagrams following Feynman's rules. A Feynman diagram can be interpreted as representing one way, that a process can take place. Importantly, the $S$ matrix element can be calculated as the sum of all topologically distinct Feynman diagrams, converted to mathematical expressions using Feynman's rules \cite{Dy49a,Dy49b,Wi50}. 

To describe electrons bound by an external electrostatic potential generated by nuclei, one can treat the continuous nucleus-electron and electron-electron interactions as infinitely many scattering events (defined above) between each pair of particles. This treatment is valid, as the lifetime of a bound state is infinitely long compared to the duration of a scattering event. This theory of bound states is quite different from the treatment of scattering free particles from before (Fig.~\ref{fig:bQED}). The key equation of this theory is the Bethe--Salpeter equation (BS equation), which defines a bound state as a pair of particles scattering infinitely many times \cite{SaBe51}. Some manipulation and reorganization of this equation yields the exact equal-time Bethe--Salpeter equation (eBS equation), or Salpeter--Sucher equation \cite{SuPhD58,MaFeJeMa23,MaMa23}.

The Salpeter--Sucher equation is the exact, relativistic wave equation for two electrons, and it is compatible with QED. By solving this equation, we are applying the most accurate physical theory to two-electron molecular systems while using the Born--Oppenheimer approximation (two electrons with fixed nuclei).\\

Historically, the golden standard of theory for comparison with precision spectroscopy in the case of light elements (low nuclear charge $Z$) has been non-relativistic QED (nrQED) \cite{Eides2001,JeAdBook22,Ye01,KoYe01,Pa06,YePaPa21,AdCaPR22}. In essence, this method consists of calculating leading-order relativistic and QED corrections to a high-precision non-relativistic reference energy ($E_0$): 
\begin{align}
  E_\text{nrQED}
  =
  E_0 
  + \alpha^2 \epsi^{(0)}_2 
  + \alpha^3 \epsi^{(0)}_3
  + \alpha^3 \ln(\alpha)\, \epsi^{(0)}_{3,1}
  + \alpha^4 \epsi^{(0)}_4
  + \alpha^4 \ln{(}\alpha{)}\; \epsi^{(0)}_{4,1}
  + \dots
  \label{eq:nrQED}
\end{align}
The non-relativistic $N$-particle Hamiltonian is treated as the zeroth-order approximation of the series expansion. Including the leading-order relativistic corrections results in the Breit--Pauli Hamiltonian, beyond which, QED corrections, \emph{e.g.,} self-energy, vacuum polarization, pair corrections can be added \cite{BeSaBook57}.
The higher-order contributions become increasingly complicated, singular (resulting in slow convergence in practical basis representations) and even divergent. The appearing divergences (beyond mass and charge renormalization) have been cancelled among the various contributions \cite{Ye01,Pa06}.

Experimentally, low-$Z$ systems are studied using precision spectroscopy. The continuous development and ever-increasing precision of experimental methods facilitate increasingly stringent testing of theoretical calculations \cite{hatom13,LiShZh91,LiShZh92,SaGi92,BeEtal98,PaGiNaHaMaIn04,PaCoGiNaInYePa12,ZhSuChJiPaHu17,ReWeNoJaEiHoVa18,ClJaScAgScMe21}. Ionization energies \cite{ClJaScAgScMe21}, and more generally, atomic transition frequencies \cite{hatom13,ZhSuChJiPaHu17,ReWeNoJaEiHoVa18} of light elements (\emph{e.g.,} H, He) can be determined with uncertainties as low as 10$^{-12}$~to~10$^{-15}$~$\Eh$.

In the case of medium to high-$Z$ elements, precision mass spectrometry measurements allow for the investigation of (atomic) properties through the energy-mass equivalence. Specifically, the measurement of electron binding energies is made possible by precise measurements of mass ratios of highly charged ions with $Z>50$ \cite{pentatrap20}.\\

It is important to add that at a lower, chemical-energy resolution, the relativistic quantum chemistry methodologies developed over the past decades $-$ of which a short and incomplete list is included in Refs.~\citenum{GoInDe87,BlJoLiSa89,PaGr90,DyFaBook07,ReWoBook15,Py12,PaElBoKaSc17,dirac20,SmInNaPiSc23} $-$
have been extremely successful in explaining the physical properties and chemical behaviour of heavy elements and their compounds. In our opinion, the power of quantum chemistry lies in having an underlying wave equation. This idea motivated the research reviewed in the present paper, \emph{i.e.,} establishing a wave equation, motivated by the success of relativistic quantum chemistry, but starting from the most possible complete and rigorous relativistic QED theory, and developing a computational framework for this theoretical formalism systematically applicable and improvable up to the ultra-high energy resolution common in precision spectroscopy and precision physics. It was not immediately obvious that such a relativistic QED wave equation exists and if it exists whether its properties allow us to use traditional basis-set type techniques. 

\begin{figure}
  \includegraphics[scale=1.]{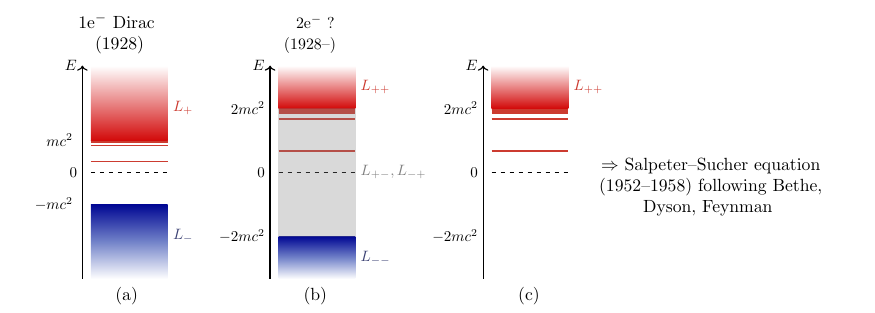}
\caption{%
  What is a fundamentally correct two-electron theory that combines special relativity and quantum mechanics? 
  Historically, the central question was: how to deal with the negative-energy states?
  In a naive two-electron generalization of the Dirac theory (b), the helium atom ground state decays with a finite lifetime~\cite{PeByKa06}.
  A correct description can be constructed with Feynman propagators and the Bethe--Salpeter equation and its equal-time variant reviewed in Sections~\ref{ch:propag}--\ref{ch:BS}.
  \label{fig:diracspectrum}
}
\end{figure}

In the following sections, we review the theoretical framework in which such a wave equation can be formulated (Fig.~\ref{fig:diracspectrum}). We also review an explicitly correlated basis set approach to solve this wave equation to high precision, and the solution can be used to systematically approach an increasingly complete relativistic QED description of atomic and molecular matter.

\section{Feynman--Stückelberg propagator theory \label{ch:propag}}
The Dirac equation for a spin-1/2 fermion with mass $m$ is
\begin{align}
  \left[%
    \iim\frac{\partial}{\partial t}-{h}(\br)
  \right]\phi(x)
  =
  0
  \quad\text{with}\quad
  h(\br)=-\iim c\balpha \bnabla_{\br} + \beta mc^2  (+ U )
  \label{eq:dirac}
\end{align}
where $x=(t,\br)$ labels the time-space coordinates with the corresponding derivatives, $\frac{\partial}{\partial t}$, $\bnabla_{\br}=(\frac{\partial}{\partial r_x},\frac{\partial}{\partial r_y},\frac{\partial}{\partial r_z})$, and $\bos{\alpha}$ and $\beta$ are the Dirac matrices.
The eigenvalue spectrum of $h$, highlighted in Fig.~\ref{fig:diracspectrum}(a), consists of both positive- and negative-energy solutions, the projectors to the corresponding $\Omega_+$ and $\Omega_-$ subspaces are labelled with $L_+$ and $L_-$, respectively. 
In the presence of an external electrostatic field, described by the $U$ potential energy contribution in Eq.~\eqref{eq:dirac}, bound states also appear in the spectrum (Fig.~\ref{fig:diracspectrum}).

For going beyond the relativistic equation of a single spin-1/2 fermion, it is better to consider the Green's function to the Dirac equation, 
\begin{align}
  \left[%
    \iim\frac{\partial}{\partial t}-{h}(\br)
  \right]g(x;x')
  =
  \iim\beta\delta(x-x')  \;. 
  \label{eq:greenSa} 
\end{align}
Multiple mathematical solutions exist depending on the combination of the positive- and negative-energy state contributions \cite{GrReBook09}, but the physically relevant choice here is according to Feynman~\cite{Fe49a}, following Stückelberg~\cite{St41}, 
\begin{align}
  &g(x;x') 
  = 
  \theta(t-t') g_+(x;x')
  - 
  \theta(t'-t) g_-(x;x')
  \label{eq:FSpropag}
  \\ 
\text{with}\quad
  &g_{\pm}(x;x')
  = 
  \sum_{k\in\Omega_{\pm}}
    \varphi_k(\br) \varphi^\dagger_k(\br') \beta \eem^{-\iim e_k (t-t')} \; , 
  \quad\text{where}\quad
  h \varphi_k(\br) = e_k \varphi_k(\br) \; .
  \label{eq:FSpropag2}  
\end{align}
The step function is $\theta(t-t')=1$ (0) for $t>t'$ ($t\leq t'$). Because of this, if $t>t'$ ($t\leq t'$), then only the $e_k>0$, positive- ($e_k<0$, negative-) energy states contribute to the propagator $g(x;x')$. This means, that the propagator $g(x;x')$ represents an electron propagating from space-time point $x'$ to $x$ if $t>t'$, while it represents a positron propagating from space-time point $x$ to $x'$ 
(as a negative-energy electron propagating from space-time point $x'$ to $x$) if $t<t'$.
It is often practical to substitute the (relative) time variable with a frequency variable. By exploiting the properties of the step function~\cite{GrReBook09}, the fermion propagator is
\begin{align}
  g(\omega)
  &= 
  \iim s(\omega) \beta\\
  s(\omega) 
  &=
  \frac{L_+}{\omega - |h| + \iim 0^+}
  + 
  \frac{L_-}{\omega + |h| - \iim 0^+} \; ,
\end{align}
where $0^+$ labels a small, positive value for which the limit is taken to zero at the end of the calculation. This construct is also known as Feynman's prescription to the fermion-propagator~\cite{Fe49a}, which is a practical mathematical tool in calculations to account for the different temporal contribution of the positive- and negative-energy one-particle Dirac states to the propagator, Eq.~\eqref{eq:FSpropag}.

\begin{figure}
  \includegraphics[scale=1.]{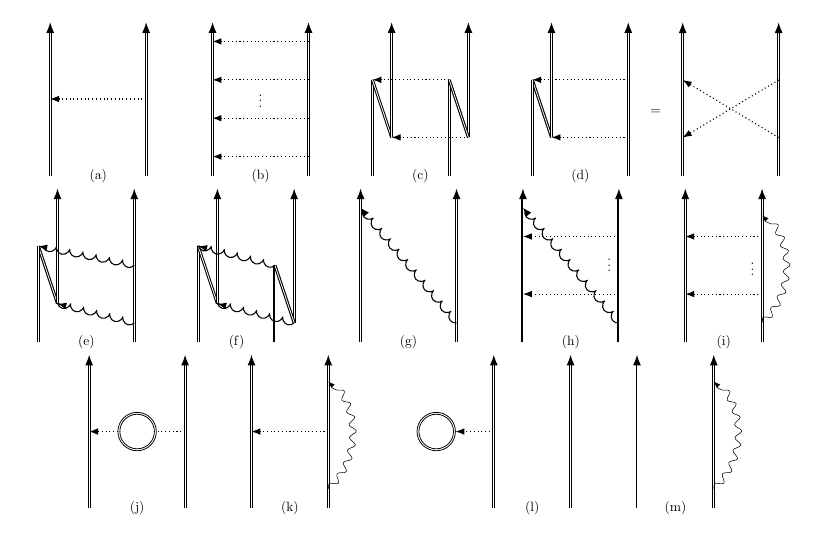}
\caption{%
  Example interaction kernels (a)--(k), vacuum polarization (l), and self energy (m). 
  Double lines indicate fermion lines in the external electrostatic field of the fixed nuclei.
  Dotted and zigzag-curvy lines are for Coulomb and transverse photons corresponding to the Coulomb gauge representation of the interaction. Wavy lines are for photons in a covariant gauge.
  Renormalization is necessary for the diagrams in the last row.
  \label{fig:kernels}
}
\end{figure}

\section{Bethe--Salpeter equation \label{ch:BS}}
The space-time propagator for two spin-1/2 fermions can be constructed according to the Bethe--Salpeter (BS) equation \cite{SaBe51} (with several formulations as well as derivations from field theory at the beginning of the 1950s \cite{Na50,GMLo51,Sc51,SaBe51})
\begin{align}
  G = g_1 g_2 - \iim g_1 g_2 K G \; ,
  \label{eq:BS}
\end{align}
where $g_a$ is the fermion propagator for particle $a$ ($a=1,2$) and $K$ collects all relevant irreducible interactions to the $G$ propagator of the interacting system.
All reducible, \emph{i.e.,} consecutive, interactions of the irreducible (non-consecutive) interactions carried by $K$ are generated by the solution of the equation.
For example, Fig.~\ref{fig:kernels}(a) shows an irreducible, whereas Figs.~\ref{fig:kernels}(b)--(c) show reducible diagrams.
The (irreducible) interaction kernels can be constructed according to Feynman's rules.

The BS equation can be renormalized according to Matthews and Salam~\cite{MaSa54}, following Dyson~\cite{Dy49b}, by using dressed and renormalized fermion propagators, photon propagators, and vertices.

For bound-state computations, the two-fermion space-time wave function satisfies the equation~\cite{GrReBook09}
\begin{align} 
  \Psi &= -\iim g_1 g_2 \tilde{K} \Psi \\
  \Psi &=  s_1 s_2 \iim \beta_1\beta_2\tilde{K} \Psi \\
  \Psi &=  s_1 s_2 \tilde{\Kcal} \Psi \; ,    
  \label{eq:BSPsi}
\end{align}
where $\tilde{K}$ collects all relevant interaction terms from $K$ and self-energy contributions from the dressed fermion-propagators \cite{MaMa23}. $\tilde{\Kcal}$ is defined for later convenience. Direct construction rules for $\tilde{\Kcal}$ are collected in the Appendix of Ref.~\citenum{MaMa23}.

In the QED theory of atoms and molecules, the binding is dominated by the instantaneous part of the electromagnetic interaction. The dominant instantaneous part is readily identified in the Coulomb gauge as the temporal part of the photon propagator, \emph{i.e.,} the Coulomb interaction. Furthermore, the instantaneous approximation to the transverse part gives rise to the Breit interaction, \emph{e.g.,} Refs.~\citenum{ReWoBook15,FeJeMa22}. 

Then, it is convenient to partition the $\tilde{\Kcal}$ interaction kernel into a $\Kcal_\inst$ instantaneous part (Coulomb or Coulomb--Breit), Fig.~\ref{fig:kernels}(a), and the rest, \emph{e.g.,} Figs.~\ref{fig:kernels}(d)--(m),
\begin{align}
  \tilde{\Kcal} = \Kcal_\inst + \Kcal_\Delta \; ,
\end{align}
where $\Kcal_\Delta$ carries all relevant irreducible contributions beyond the $\Kcal_\inst$ instantaneous interaction.

This partitioning motivates further rearrangements,
\begin{align}
  s^{-1}_1 s^{-1}_2 |\Psi\rangle
  &= 
  (\Kcal_\inst + \Kcal_\Delta) |\Psi\rangle \\
  \left[%
    s^{-1}_1 s^{-1}_2 - \Kcal_\Delta
  \right]
  |\Psi \rangle
  &= 
  \Kcal_\inst |\Psi\rangle \\
  |\Psi\rangle
  &= 
  \left[%
    s^{-1}_1 s^{-1}_2 - \Kcal_\Delta
  \right]^{-1}
  \Kcal_\inst |\Psi\rangle \\ 
  |\Psi\rangle
  &=
  \left[%
  s_1 s_2
  +
  s_1 s_2 \Kcal_\Delta
  \left(%
    s^{-1}_1 s^{-1}_2 - \Kcal_\Delta
  \right)^{-1}
  \right]
  \Kcal_\inst |\Psi\rangle \\ 
  |\psi\rangle
  &=
  \left[%
  s_1 s_2
  +
  s_1 s_2 \Kcal_\Delta
  \left(%
    1 - s_1 s_2 \Kcal_\Delta
  \right)^{-1} s_1 s_2
  \right]
  \Kcal_\inst |\psi\rangle 
\end{align}
where in the last step, we factorized and cancelled the absolute time, $T=(t_1+t_2)/2$ dependence by using $|\Psi\rangle = |\psi\rangle\eem^{-\iim E T}$ and exploiting the fact that any external field is time-independent in our systems. 
Then, the $s_1$, $s_2$ frequency (energy) dependence simplifies to (notation details are explained in Ref.~\citenum{MaMa23})
\begin{align}
  s_1(\epsi) 
  &= 
  \frac{%
    1
  }{%
  \frac{E}{2}+\epsi - h_1 + \iim 0^+ (L_{1+}-L_{1-})
  } \\
  s_2(-\epsi) 
  &= 
  \frac{%
    1
  }{%
  \frac{E}{2}-\epsi - h_2 + \iim 0^+ (L_{2+}-L_{2-})
  } \; .
\end{align}
We proceed by integrating out the $\epsi$ (half of the) relative energy, which is the Fourier pair of the $t_1-t_2$ relative time dependence, and obtain a solely space- (or momentum-) dependent `equal-time' wave function as
\begin{align}
  |\Phi\rangle 
  &
  :=\int_{-\infty}^{+\infty} \dd\epsi\ 
  |\psi(\epsi)\rangle 
  \nonumber \\
  &=
  \int_{-\infty}^{+\infty} \dd\epsi
  \left\lbrace%
  s_1 s_2
  +
  s_1 s_2 \Kcal_\Delta
  \left[%
    1 - s_1 s_2 \Kcal_\Delta
  \right]^{-1}
  s_1 s_2
  \right\rbrace
  \frac{1}{-2\pi\iim} V_\inst \int_{-\infty}^{+\infty}\dd\epsi'\ |\psi(\epsi')\rangle \\   
  &=
  \int_{-\infty}^{+\infty} \frac{\dd\epsi}{-2\pi\iim}
    s_1 s_2 V_\inst |\Phi\rangle 
  +
  \int_{-\infty}^{+\infty} \frac{\dd\epsi}{-2\pi\iim}  
    s_1 s_2 \Kcal_\Delta
    \left[%
      1 - s_1 s_2 \Kcal_\Delta
    \right]^{-1}
    s_1 s_2
    V_\inst |\Phi\rangle \\
  &=
  \frac{L_{++}-L_{--}}{E-h_1-h_2}
   V_\inst |\Phi\rangle 
  +
  \int_{-\infty}^{+\infty} \frac{\dd\epsi}{-2\pi\iim}  
    s_1 s_2 \Kcal_\Delta
    \left[%
      1 - s_1 s_2 \Kcal_\Delta
    \right]^{-1}
    s_1 s_2
    V_\inst |\Phi\rangle  \; .
  \label{eq:phi}
\end{align}
During the course of the calculations, we exploited the `trivial' relative-energy dependence of the instantaneous interaction, which results in \cite{MaMa23}
\begin{align}
  \Kcal_\inst |\psi(\epsi)\rangle 
  =
  \frac{1}{-2\pi\iim} V_\inst |\Phi\rangle \; .
\end{align}
Furthermore, the relative-energy integral of single-particle propagator products is
\begin{align}
  \int_{-\infty}^{+\infty}\frac{\dd \epsi}{-2\pi\iim}
    s_1(\epsi)s_2(-\epsi)
  =
  \frac{L_{++}-L_{--}}{E-h_1-h_2} \; ,
\end{align}
which can be calculated using Cauchy's residue theorem or the Sokhotski--Plemelj formula \cite{MaMa23}. 
Further details and calculation techniques can be found in Refs.~\citenum{MaFeJeMa23,MaMa23}. In the two-particle theory, we use the short notation for two-particle projectors, $L_{\sigma_1\sigma_2}=L_{1\sigma_1}L_{2\sigma_2}\ (\sigma_1,\sigma_2=+,-)$, \emph{e.g.,} $L_{++}=L_{1+}L_{2+}$.

The rearrangement of Eq.~\eqref{eq:phi} results in a non-linear eigenvalue(-like) equation, 
the equal-time Bethe--Salpeter equation, first formulated by Salpeter \cite{Sa52} and Sucher \cite{SuPhD58}, 
\begin{align}
  E|\Phi\rangle 
  =
  \left[
    h_1 + h_2 + (L_{++}-L_{--}) V_\inst + \Vcal_\epsilon(E)
  \right]|\Phi\rangle \; .
  \label{eq:eBS1}
\end{align}
This is the exact QED wave equation for the $\Phi$ equal-time wave function, which depends only on the coordinates (momenta) of the two particles and has no explicit (relative) time (relative energy) dependence. The $\epsi$ relative-energy dependence is carried by the $\Vcal_\epsilon(E)$ `potential-energy-like' term, 
\begin{align}
  \Vcal_\epsilon(E)
  =
  (E-h_1-h_2)
  \int_{-\infty}^{+\infty} \frac{\dd\epsi}{-2\pi\iim}  
    s_1 s_2 \Kcal_\Delta
    \left[%
      1 - s_1 s_2 \Kcal_\Delta
    \right]^{-1}
    s_1 s_2
    V_\inst \; ,
    \label{eq:Vcal}
\end{align}
in which the $s_1,s_2$ and other factors in $\Kcal_\Delta$ depend on the $E$ total energy, rendering Eq.~\eqref{eq:eBS1} nonlinear in $E$. $\Vcal_\epsilon$ carries the effect of (irreducible) pair, retardation, and radiative corrections, Figs.~\ref{fig:kernels}(d)--(m). In QED, their energetic contribution can be anticipated to be small, in comparison with the instantaneous interaction ladder, Figs.~\ref{fig:kernels}(a)--(b). Furthermore, the double pair-correction, Fig.~\ref{fig:kernels}(c), due to the instantaneous interaction ($V_\delta$, \emph{vide infra}) is also small; hence, a natural partitioning of Eq.~\eqref{eq:eBS1} is obtained as
\begin{align}
  \left[
  H_\nopair
  +
  V_\delta + \Vcal_\epsilon(E)
  \right]
  |\Phi\rangle
  =
  E |\Phi\rangle \; ,
  \label{eq:eBS2}
\end{align}
where $H_\nopair$ is the (hermitian) no-pair Dirac--Coulomb or Dirac--Coulomb--Breit (DC or DCB) Hamiltonian,
\begin{align}
    H_\nopair
    =
    h_1 + h_2
    +
    L_{++} V_\inst L_{++} \; ,
    \label{eq:npDCB}
\end{align}
and the non-crossing pair correction due to the instantaneous interaction, 
\begin{align}
  V_\delta
  =
  L_{++} V_\inst (1-L_{++})
  -
  L_{--} V_\inst \; .
  \label{eq:Vdelta}
\end{align}

All in all, it can be anticipated that we obtain a good (starting) approximation by considering only the linear and hermitian part of the full operator in Eq.~\eqref{eq:eBS2}, and first solve the no-pair DC(B) eigenvalue equation,
\begin{align}
    H_\nopair
    |\Phi_\nopair\rangle
    &=
    E_\nopair |\Phi_\nopair\rangle \; .
    \label{eq:Hnopair}
\end{align}
We also note that the `non-trivial' interacting part of this equation belongs to the positive-energy subspace ($L_{++}$), which is not coupled to the $+-$ and $-+$ (Brown--Ravenhall, BR), and $--$ subspaces by $H_\nopair$. Hence, it is convenient to define
\begin{align}\label{eq:HppEigen}
    H^\pp
    |\Phi_\nopair\rangle
    &=
    E_\nopair |\Phi_\nopair\rangle \; 
\end{align}
with
\begin{align}\label{eq:Hpp1}
    H^\pp
    &=
    L_{++} H_\nopair L_{++}
    \\\label{eq:Hpp2}
    &=
    L_{++} \left[
        h_1 + h_2
        +
        V_\inst
    \right] L_{++} \; .
\end{align}
Since this Hamiltonian is hermitian and bounded from below, it can be solved by variational, energy minimization techniques, \emph{e.g.,} Ref.~\citenum{SuVaBook98}, to high precision \cite{JeFeMa21,JeFeMa22,FeJeMa22,FeJeMa22b,JeMa23,FeMa23}. Concerning the no-pair energy (Tables~\ref{tab:He1S_DCB}--\ref{tab:He2S_DCB}), we will always refer to this non-trivial $++$ part and highlight an explicitly correlated variational solution approach recently developed for its computation.
Then, the last part of the paper outlines plans for a perturbative account of the energy-dependent $\VQED(E)=V_\delta + \Vcal_\epsilon(E)$ term \citep{MaFeJeMa23,MaMa23}, in which, of course, some of the operators will act on the $+-$, $-+$, and $--$ subspaces in a non-trivial manner.

\begin{figure}
\hspace{-2.0cm}%
\includegraphics[scale=0.7]{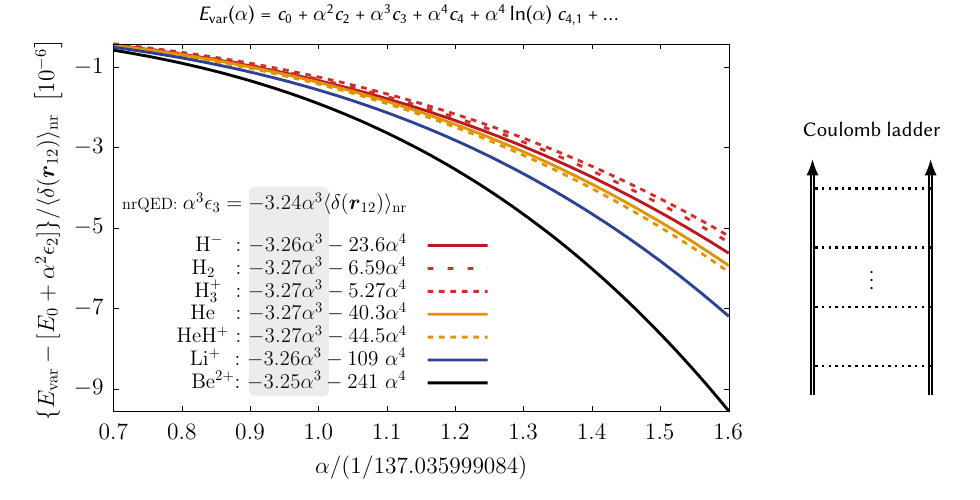}
\caption{%
The $\alpha$ fine-structure-constant dependence of the high-precision variational no-pair Dirac--Coulomb(--Breit) energy is consistent with the most extensively tested perturbative `nrQED' theory based on a high-precision non-relativistic reference.
The variational computation automatically includes the full instantaneous Coulomb(--Breit) interaction ladder in the relativistic treatment. The dataset highlighted in the figure is taken from Ref.~\citenum{JeFeMa22}. The higher-order $\alpha$ dependence is revealed by high-precision computations detailed in Table~\ref{tab:alphascaling} (see also Ref.~\citenum{FeMa23}), and it is in agreement with higher-order nrQED correction values.
}
\label{fig:alphascal}
\end{figure}

\section{Explicitly correlated variational solution of the no-pair Dirac--Coulomb(--Breit) equation \label{ch:numres}}

\subsection{Theoretical aspects of the numerical implementation}

For the variational solution of the no-pair DC(B) equation, we solve the eigenvalue equation of the fully positive-energy projected DC(B) Hamiltonian, Eqs.~(\ref{eq:HppEigen})--(\ref{eq:Hpp2}), 
\begin{align}
    H^\pp
    =
    L_{++} \qty[
        h\four_1 \boxtimes 1\four
        +
        1\four \boxtimes h\four_2
        +
        V_\tC 1\sixteen
        \qty(+ V\sixteen_\tB)
    ] L_{++}
    \; ,
\end{align}
where the instantaneous interaction $V_\inst$ can be either the Coulomb interaction, $V_\tC$, or the Coulomb--Breit interaction, $V_\tCB=V_\tC+V_\tB$ \cite{MaMa23}, and $\boxtimes$ is the block-wise direct product. The explicit form of this Hamiltonian is 
\begin{align}
    &H^\pp
    =
    \nonumber\\
    &
    L_{++}
    \mqty(
        \qty(U + V_\tC) 1\four &
        c\bsigma\four_2 \bp_2 &
        c\bsigma\four_1 \bp_1 &
        V\four_\tB \\
        c\bsigma\four_2 \bp_2 &
        \qty(U + V_\tC - 2m_2c^2) 1\four &
        V\four_\tB &
        c\bsigma\four_1 \bp_1 \\
        c\bsigma\four_1 \bp_1 &
        V\four_\tB &
        \qty(U + V_\tC - 2m_1c^2) 1\four &
        c\bsigma\four_2 \bp_2 \\
        V\four_\tB &
        c\bsigma\four_1 \bp_1 &
        c\bsigma\four_2 \bp_2 &
        \qty(U + V_\tC - 2m_{12}c^2) 1\four
    )
    L_{++}
\end{align}
with $m_{12}=m_1+m_2$, and with the energy scale shifted by the rest energies of the two electrons. This shift serves the practical purpose of making the numerical results readily comparable to the non-relativistic energy. Also, $\bsigma\four_1 = \bsigma\ntwo \otimes 1\ntwo$ and $\bsigma\four_2 = 1\ntwo \otimes \bsigma\ntwo$, where $\bsigma\ntwo = (\si\ntwo_1,\si\ntwo_2,\si\ntwo_3)$ are the Pauli matrices.

The Ansatz for the sixteen-component eigenspinor of the no-pair DC(B) Hamiltonian is \cite{JeFeMa22,FeJeMa22,FeJeMa22b}
\begin{align}
    |\Phi_\nopair\rangle
    =
    \Acal\sixteen
    \sum_{q=1}^{16}
    \sum_{\mu=1}^N
    c_{q \mu} f_\mu
    X\sixteen |1\sixteen_q\rangle
\end{align}
with the spatial part of the spinor expressed with a floating explicitly correlated Gaussian (fECG) basis set, which provides here a two-electron basis including electron correlation. An fECG basis function, in general, takes the form
\begin{align}
    f_\mu
    =
    \exp\qty[
        -{(\br-\bs_\mu)}^\tT\buA_\mu(\br-\bs_\mu)
    ]
    \; ,
    \label{eq:ecg}
\end{align}
where $\br=(\br_1,\br_2)$ contains the spatial coordinates of both electrons, $\bs_\mu$ is the shift vector defining the position where the basis function is centered, and $\buA_\mu=\bA\ntwo_\mu\otimes1\nthree$ with $\bA\ntwo_\mu$ being a symmetric positive definite matrix containing the basis function parameters.

The large and small components of the spinor are balanced using the simplest two-particle generalization of the `restricted' kinetic balance for a one-particle spinor \cite{JeFeMa22}. This generalization is
\begin{align}
    X\sixteen
    =
    X\four_1 \boxtimes X\four_2
    =
    \mqty[
        1\four &
        0\four &
        0\four &
        0\four \\
        0\four &
        \frac{\bsigma\four_2 \bp_2}{2m_2c} &
        0\four &
        0\four \\
        0\four &
        0\four &
        \frac{\bsigma\four_1 \bp_1}{2m_1c} &
        0\four \\
        0\four &
        0\four &
        0\four &
        \frac{(\bsigma\four_1 \bp_1)(\bsigma\four_2 \bp_2)}{4m_1m_2c^2} 
    ]
\end{align}
with $X\four_i$ being the `restricted' kinetic balance for the 4-spinor of the $i$-th particle.

Finally, $\Acal\sixteen=\frac{1}{2}(1\sixteen-\Pcal\sixteen_{12})$ is the antisymmetrization operator defined in Ref.~\citenum{JeFeMa22}, such that the permutation operator $\Pcal\sixteen_{12}$ exchanges the particle indices and the corresponding spinor structure.

To obtain the projection operator $L_{++}$, we are required to solve the eigenvalue equation of the non-interacting two-particle Hamiltonian, $h_1+h_2$. For low-$Z$ systems, the `cutting' projection operators were found to perform remarkably well. These are defined as \cite{JeFeMa22}
\begin{align}
    L_{++}
    =
    \sum_{i\in\Omega_{++}}
    |\Phi^{(0)}_i\rangle\langle\Phi^{(0)}_i|
\end{align}
with
\begin{align}
    |\Phi^{(0)}_i\rangle
    =
    \Acal\sixteen
    \sum_{q=1}^{16}
    \sum_{\mu=1}^N
    c^{(0)}_{i, q \mu} f_\mu
    X\sixteen |1\sixteen_q\rangle
    \qq{where}
    \qty[h_1 + h_2]
    |\Phi^{(0)}_i\rangle
    =
    E^{(0)}_i
    |\Phi^{(0)}_i\rangle
    \; ,
\end{align}
and thus, only those states are retained which have their energy above a threshold. This threshold energy is chosen as a lower-bound to the non-interacting ground-state energy.
This cutting projector provides an approximate separation of the ++ and the higher-energy part of the BR space, but it is demonstrated to be an extremely good approximation to the rigorous $L_{1+}L_{2+}$ `one-particle' projection approach using one-particle operator representations over a two-particle basis space (Fig.~\ref{fig:permsym}) \cite{HoJeMa24}. 

All computations have been carried out with the in-house developed computer program named QUANTEN, which contains an extensive collection of ECG integrals, and a non-linear variational engine, and has been recently extended with OpenMP parallel increased-precision linear algebra subroutines.

\subsection{High-precision correlated, relativistic energy for the example of the helium atom}
By increased-precision computations, we could extend earlier work \cite{JeFeMa22,FeJeMa22,FeJeMa22b} towards the part-per-trillion (ppt) relative precision range of the no-pair Dirac--Coulomb(--Breit) energy. 
Similarly to earlier work, ECG spatial functions were parameterized by minimization of the non-relativistic energy.
In the largest computations, the 1S (1~$^1$S$_0$) and 2S (2~$^1$S$_0$) non-relativistic energies of the helium atom were converged to 38 and 102~p$\Eh$ (p$\Eh=10^{-12}\ \Eh$), respectively, in comparison with the benchmark non-relativistic energy of Refs.~\citenum{Dr06,DrYa92,AzBeKo18}.
This improved parameterization allowed us to converge further digits of the no-pair DC(B) energy, Table~\ref{tab:He1S_DCB} and \ref{tab:He2S_DCB} (and \som), in the singlet basis sector. The triplet basis functions (of $\alpha^4\Eh$ order in nrQED) can be added to the basis space according to Ref.~\citenum{JeMa24}.
The improved parameterization reported in the present work, and the better-converged results allow us to pinpoint the $\alpha^4$ order coefficient, and we can also determine the coefficient of the $\alpha^4\ln{(}\alpha{)}$ term of the 1S state, in excellent agreement (Table~\ref{tab:alphascaling}) with nrQED calculations \cite{KhMiYe93,DrKhMiYe93,Zh96}.

The $\alpha^4\ln{(}\alpha{)}$-order term has already been obtained in pseudo-one-particle, sixteen-component pre-Born--Oppenheimer Dirac--Coulomb(--Breit) computations for positronium-like systems \cite{FeMa23}, in which the small basis sizes (30--50 functions) could be handled with serial (single core) quadruple-precision linear-algebra subroutines. 

At the present level of convergence of the helium atom energies (Tables~\ref{tab:He1S_DCB} and \ref{tab:He2S_DCB}), signs of the $\alpha^5$ dependence are also seen. These numerical results already provide a solid starting point for a perturbation theory treatment of the $\Vcal_\epsilon$ term, Eq.~\eqref{eq:Vcal}, for the kernels collected in Fig.~\ref{fig:kernels}, with respect to the correlated relativistic reference state. Furthermore, we plan to continue our computations to pinpoint the no-pair Dirac--Coulomb and Dirac--Coulomb--Breit energies to (sub)-ppt relative precision, and to determine the higher-order $\alpha$ dependence.

\begin{table}
    \centering
    \caption{He 1S no-pair DC(B) energy, in $\Eh$.}
    \label{tab:He1S_DCB}
    \begin{tabular}{@{}l@{}c ll@{}}
	\hline \hline
        Prec.$^\text{a}$ & 
	$N$$^\text{b}$ & 
        \multicolumn{1}{c}{$E^{\pupu}_\DC$} & 
        \multicolumn{1}{c}{$E^{\pupu}_\DCB$} \\
        \hline
        D$^\text{a}$  &
        700             & --2.903 856 631 6 
                        & --2.903 828 121 1  \\
        \hline
        \multirow{6}{*}{Q$^\text{a}$}
        &500            & --2.903 856 632 09
                        & --2.903 828 122 62\\
        &600            & --2.903 856 632 171
                        & --2.903 828 122 815\\
        &750            & --2.903 856 632 202
                        & --2.903 828 122 865\\
        &1000           & --2.903 856 632 211
                        & --2.903 828 122 878\\
        &1000r          & --2.903 856 632 233
                        & --2.903 828 122 927\\
        &1000rr         & --2.903 856 632 240
                        & --2.903 828 122 937\\
        \hline \hline
    \end{tabular}
    \flushleft{\footnotesize{%
      $^\text{a}$~Computations using 8-byte (D) \cite{JeFeMa22,FeJeMa22b} and 16-byte (Q) reals (this work).  \\
      $^\text{b}$~Number of ECGs optimized to the non-relativistic energy, `r' and `rr' means further basis optimization cycles. The best non-relativistic energy is converged within 38~p$\Eh$ \cite{Dr06,AzBeKo18}. 
      Only the singlet basis sector is shown; the triplet basis space, within the $LS$ coupling scheme, can be added to the basis set according to Ref.~\citenum{JeMa23}. 
    }}
\end{table}

\begin{table}
    \centering
    \caption{He 2S no-pair DC(B) energy, in $\Eh$.}
    \label{tab:He2S_DCB}
    \begin{tabular}{@{}l@{}c ll@{}}
	\hline        \hline
        Prec. & 
	$N$ & 
        \multicolumn{1}{c}{$E^{\pupu}_\DC$} & 
        \multicolumn{1}{c}{$E^{\pupu}_\DCB$} \\
        \hline
        D &
        400             & --2.146 084 791 
                        & --2.146 082 363 \\    
        \hline
        \multirow{4}{*}{Q}
        &500            & --2.146 084 791 450
                        & --2.146 082 363 473 \\
        &600            & --2.146 084 791 467
                        & --2.146 082 363 502 \\
        &800            & --2.146 084 791 480
                        & --2.146 082 363 523 \\
        &1000           & --2.146 084 791 490
                        & --2.146 082 363 552 \\
        \hline \hline
    \end{tabular}
    \flushleft{\footnotesize{
       See the footnotes to Table~\ref{tab:He1S_DCB}. The best non-relativistic energy is converged within 102~p$\Eh$\ with respect to Ref.~\citenum{AzBeKo18}. 
    }}
\end{table}

\begin{table}
  \caption{%
    Variational no-pair Dirac--Coulomb(--Breit) energies (Var) of the helium atom 1S and 2S states compared, through their $\alpha$ dependence $E_\text{var}(\alpha)=c_0+c_2\alpha^2 + c_3\alpha^3 + c_4\alpha^4 [+ c_{4,1}\alpha^4 \ln{(}\alpha{)}]$,  with terms of the nrQED expansion, Eq.~\eqref{eq:nrQED}, compiled from the literature  \cite{SuPhD58,Pa06,KhMiYe93,DrYa92,Dr06,AzBeKo18}. 
    \label{tab:alphascaling}
  }
  \scalebox{0.9}{%
  \begin{tabular}{@{}ll lll ll@{}}
    \hline\hline  
    & &
    \multicolumn{1}{c}{$c_0$} & 
    \multicolumn{1}{c}{$c_2$} & 
    \multicolumn{1}{c}{$c_3$} & 
    \multicolumn{1}{c}{$c_4$} & 
    \multicolumn{1}{c}{$c_{4,1}$} \\
    \hline \\[-0.3cm] 
    \multicolumn{7}{l}{\textbf{He 1S, Dirac--Coulomb Hamiltonian: 
    }} \\
    \multicolumn{2}{l}{nrQED}
    & --2.903 724 377 034
    & --2.480 847 8
    & --0.344 29
    & [--3.030]$^\text{d}$
    & 0.167$^\text{e}$ \\
    4:   & Var$^\text{a}$ 
    & --2.903 724 377 022
    & --2.480 839 8
    & --0.347 26
    & --4.301
    & --
    \\
    4,1: & Var$^\text{a}$ 
    & --2.903 724 376 996
    & --2.480 846 7
    & --0.344 53
    & --3.822
    &   0.149
    \\
    4:   & $(\delta\epsi_n)\alpha^n[\text{n}\Eh]$$^\text{b}$
    & \multicolumn{1}{r}{--0.012}
    & \multicolumn{1}{r}{--0.426}
    & \multicolumn{1}{r}{1.156}
    & \multicolumn{1}{r}{[3.604]$^\text{d}$}
    & \multicolumn{1}{r}{--}
    \\
    4,1: & $(\delta\epsi_n)\alpha^n[\text{n}\Eh]$$^\text{b}$
    & \multicolumn{1}{r}{--0.038}
    & \multicolumn{1}{r}{--0.055}
    & \multicolumn{1}{r}{0.095}
    & \multicolumn{1}{r}{[2.245]$^\text{d}$}
    & \multicolumn{1}{r}{--0.254}
    \\\hline\\[-0.2cm]
    \multicolumn{7}{l}{\textbf{He 2S, Dirac--Coulomb Hamiltonian:}} \\
    \multicolumn{2}{l}{nrQED}
    & --2.145 974 046 054
    & --2.079 252 9 
    & --0.028 00 
    & [--3.970]$^\text{d}$
    & 0.014$^\text{e}$ 
    \\  
    4:   & Var$^\text{a}$ 
    & --2.145 974 045 952
    & --2.079 253 4
    & --0.028 16
    & --4.068
    & --
    \\
    4,1: & Var$^\text{a}$ 
    & --2.145 974 045 951 
    & --2.079 253 5
    & --0.028 13
    & --4.063 
    &   0.002
    \\      
    4:  & $(\delta\epsi_n)\alpha^n[\text{n}\Eh]$$^\text{b}$
    & \multicolumn{1}{r}{--0.102}
    & \multicolumn{1}{r}{0.028} 
    & \multicolumn{1}{r}{0.063}
    & \multicolumn{1}{r}{[0.278]$^\text{d}$}
    & \multicolumn{1}{r}{--}
    \\    
    4,1: & $(\delta\epsi_n)\alpha^n[\text{n}\Eh]$$^\text{b}$
    & \multicolumn{1}{r}{--0.103}
    & \multicolumn{1}{r}{0.032}
    & \multicolumn{1}{r}{0.051}
    & \multicolumn{1}{r}{[0.263]$^\text{d}$}
    & \multicolumn{1}{r}{--0.165}
    \\\hline\\[-0.2cm]
    %
    \multicolumn{7}{l}{\textbf{He 1S, Dirac--Coulomb--Breit Hamiltonian:}} \\
    \multicolumn{2}{l}{nrQED}
    & --2.903 724 377 034
    & --1.951 754 8
    & [0.582 23]$^\text{c}$
    & [--3.030]$^\text{d}$ 
    & 
    \\
    4:   & Var$^\text{a}$ 
    & --2.903 724 376 893
    & --1.951 774 6
    & 0.515 17
    & --4.088
    & --
    \\
    4,1: & Var$^\text{a}$ 
    & --2.903 724 376 985
    & --1.951 749 8
    & 0.505 42
    & --5.800
    & --0.531
    \\    
    4:   & $(\delta\epsi_n)\alpha^n[\text{n}\Eh]$$^\text{b}$
    & \multicolumn{1}{r}{--0.141}
    & \multicolumn{1}{r}{1.057}
    & \multicolumn{1}{r}{[26.1]$^\text{c}$}
    & \multicolumn{1}{r}{[3.000]$^\text{d}$}
    & \multicolumn{1}{r}{--}
    \\  
    4,1: & $(\delta\epsi_n)\alpha^n[\text{n}\Eh]$$^\text{b}$
    & \multicolumn{1}{r}{--0.049}
    & \multicolumn{1}{r}{--0.094}
    & \multicolumn{1}{r}{[29.8]$^\text{c}$}
    & \multicolumn{1}{r}{[7.855]$^\text{d}$}
    & 
    \\\hline\\[-0.2cm]
    \multicolumn{7}{l}{\textbf{He 2S, Dirac--Coulomb--Breit Hamiltonian:}} \\
    \multicolumn{2}{l}{nrQED} 
    & --2.145 974 046 054
    & --2.034 168 9
    & [0.047 35]$^\text{c}$ 
    & [--3.970]$^\text{d}$
    \\  
    4:    & Var$^\text{a}$ 
    & --2.145 974 045 935 
    & --2.034 166 8
    & 0.041 77
    & --4.129
    & --
    \\
    4,1: & Var$^\text{a}$ 
    & --2.145 974 045 946
    & --2.034 163 8
    & 0.040 59
    & --4.335
    & --0.064
    \\    
    4:   & $(\delta\epsi_n)\alpha^n[\text{n}\Eh]$$^\text{b}$
    & \multicolumn{1}{r}{--0.120}
    & \multicolumn{1}{r}{--0.109}
    & \multicolumn{1}{r}{[2.169]$^\text{c}$}
    & \multicolumn{1}{r}{[0.452]$^\text{d}$}
    & \multicolumn{1}{r}{--}
    \\    
    4,1: & $(\delta\epsi_n)\alpha^n[\text{n}\Eh]$$^\text{b}$
    & \multicolumn{1}{r}{--0.108}
    & \multicolumn{1}{r}{--0.269}
    & \multicolumn{1}{r}{[2.625]$^\text{c}$}
    & \multicolumn{1}{r}{[1.036]$^\text{d}$}
    & 
    \\        
    \hline\hline
  \end{tabular}
}      
  \begin{flushleft}
  $^\text{a}$~%
    The $c_0,c_2,c_3,c_4$ (and $c_{4,1}$) coefficients were computed by solving a system of linear equations 
    for $E_\text{var}(\alpha)$ with the $\alpha$ fine-structure constant tuned to $\alpha^{-1}-\alpha_0^{-1}=$ 50, 25, 0, --25, (--50) with $\alpha^{-1}_0=137.035999084$ being the CODATA18 recommended value.
    In all computations, the non-linear basis parameterization, Eq.~\eqref{eq:ecg}, was taken   from the minimization of the non-relativistic energy converged to 38~p$\Eh$\ (102~p$\Eh$) for the 1S (2S) state \cite{Dr06,AzBeKo18}. \\ 
  $^\text{b}$~%
    $(\delta\epsi_n)\alpha^n$, in n$E_\text{h}$, with the $\delta\epsi_0=E_0-c_0$  and $\delta\epsi_n = \epsi^{\nulla}_n-c_n\ (n>0)$ difference of the nrQED value (known to more digits than shown in the Table), Eq.~\eqref{eq:nrQED}, and the fitted coefficient. For the logarithmic term, 
    the $(\epsi^{\nulla}_{4,1}-c_{4,1})\alpha^4\ln{(}\alpha{)}$ deviation is reported. \\
  $^\text{c}$~No direct comparison is possible. 
  The currently available nrQED  value \cite{SuPhD58} corresponds to 
$\epsi_3^{\nulla}=\epsi_{\text{C$\cdot$C}\pupu}+\epsi_{\text{C$\cdot$B}\pupu}+\epsi_{\text{T$\cdot$T}\pupu}=$ $\left(\pi+\frac{7}{3}\right)\langle\delta(\br_{12})\rangle_\text{nr}$, where $\epsi_{\text{T$\cdot$T}\pupu}$ includes also retardation effects in the double-transverse-photon exchange \cite{SuPhD58}, which can be estimated as the difference of this $\epsi_3^{\nulla}$ and the $c_3^\text{(DCB)}$ value to be 29.8 (2.6) n$\Eh$ for the 1S (2S) state.  \\
  $^\text{d}$~No direct comparison is possible. The currently available nrQED value of Ref.~\citenum{Pa06}, `$(E_A'+E_Q)\alpha^4$', includes also `further' QED (pair and retardation) corrections. \\
  $^\text{e}$ $\epsi^{(0)}_{4,1}=\frac{\pi}{2} \langle \delta(\br_{12})\rangle_\text{nr}$ based on Refs.~\cite{KhMiYe93,DrKhMiYe93,Zh96}. \\
  \end{flushleft}
\end{table}

\section{Development of a \emph{relativistic} QED perturbation theory \label{ch:vqed}}
\noindent %
The no-pair DC(B) energy is orders of magnitude more accurate than the non-relativistic energy (Table~\ref{tab:nrQEDcorr} for $Z=2$), hence it can provide a better starting point for a perturbation theory treatment, which we call \emph{relativistic} QED (rQED) in comparison with the tranditional \emph{non-relativistic} QED (nrQED) approach. Furthermore, as far as it can be seen at the moment, rQED has a mathematically simpler structure than nrQED, and most importantly, at intermediate computational steps, it suffers from far fewer divergences than nrQED. In particular, rQED appears to have `only' the `inherent' QED divergences corresponding to mass and charge renormalization. 

All in all, relying on the high-precision no-pair energies and wave functions (Tables~\ref{tab:He1S_DCB}--\ref{tab:alphascaling}, Fig.~\ref{fig:alphascal}), it is highly relevant to compute perturbation theory corrections for the retardation-, pair-, and crossed-photon processes collected in Fig.~\ref{fig:kernels}.
The orders of magnitude of the contributions, and hence, their `importance', can be estimated based on the nrQED expressions of Sucher \cite{SuPhD58}, which sum to the well-known $\alpha^3\Eh$-order, so-called `leading-order' QED correction~\cite{Ar57,SuPhD58} with many applications in precision spectroscopy computations of (two- and few-electron) low-$Z$ systems, \emph{e.g.,} Refs.~\citenum{PiJe09,KoPiLaPrJePa11,PrCeKoLaJeSz10,FeKoMa20,FeMa19EF,FeMa19HH,FeMa23bethe,SaFeMa23,LeLa23}.

\begin{table}
\caption{%
  $\alpha^3\Eh$-order nrQED correction to the non-relativistic singlet energy of the ground-state helium atom compiled from Refs.~\citenum{SuPhD58,Ar57} (see also Fig.~\ref{fig:kernels}). For comparison, the $\alpha^2\Eh$, leading-order relativistic correction is $\alpha^2\epsi_2^\nulla=-103\ 933.59\ \text{n}\Eh$ \cite{FeJeMa22b,Dr06}, and it is automatically included in the no-pair DC(B) energy.
  \label{tab:nrQEDcorr}
}
\scalebox{0.8}{%
\begin{tabular}{@{}l@{}r@{} c@{}r@{}r@{}}
\hline\hline\\[-0.30cm]
  & $\delta\epsi^\nulla_3$
  & \citenum{Ar57,SuPhD58}$^\text{a}$
  & \multicolumn{2}{c}{$\delta\epsi^{\nulla}_3\alpha^3$ [n$\Eh$]$^\text{b}$ } \\
  \hline\\[-0.25cm]
  \textbf{Double-Coulomb-photon correction}
  & $-\frac{4}{3}\delta_{12}$ 
  & (Ar7.4-5)
  & 
  & \textbf{--55.10} \\
  
  incl. C$\cdot$C$\pupu$
  & $-\left(\frac{\pi}{2}+\frac{5}{3}\right) \delta_{12}$ 
  & (Su3.99)
  & --133.79$^\text{c}$ %
  &  \\
  incl. C$\times$C$\pumi$
  & $2 \delta_{12}$ 
  & (Su4.26a)
  & 82.65
  & \\
  incl. C$\cdot$C$\mimi$ 
  & $\left(\frac{\pi}{2}-\frac{5}{3} \right) \delta_{12}$ 
  & (Su4.26b)
  & --3.96 
  & \\
  \hline\\[-0.30cm] 
  \textbf{Double-transverse-photon correction}
  & $[\frac{14}{3} - \frac{8}{3}\ln 2 + 2\ln\alpha ]\delta_{12} - 2Q$
  & (Ar5.11)
  & 
  & \textbf{--351.38}  \\
  incl. T$\cdot$T$\pupu$   
  & $-\frac{\pi}{2} \delta_{12}$
  & (Su6.9b)
  & --64.91$^\text{c}$
  &  \\ 
  incl. T$\times$T$\pumi$
  & $4\ln 2 \delta_{12}$
  & (Su6.9b) 
  & 114.58
  & \\
  incl. T$\cdot$T$\mimi$ (regular part)
  & $[(\frac{\pi}{2}-2\ln 2)\delta_{12}]$
  & (Su6.9b)
  & 7.62 
  & \\
  \hline \\[-0.25cm]
  \textbf{Transverse, self-energy, vertex corr.}
  & 
  \multicolumn{2}{r}{[(Ar3.26)$-w(0,\infty)$]+(Ar6.12)}
  & 
  & \textbf{--6629.35} \\ 
  incl. `Bethe logarithm' contribution
  & $-\frac{4}{3} \text{ln}k_0 Z(\delta_1+\delta_2)\alpha^3$ 
  & \cite{SuPhD58,Ar57}
  & --16\ 397.35
  & \\  
  incl. part of the `Araki--Sucher' term 
  & $-\frac{8}{3}Q$
  & \cite{SuPhD58,Ar57}
  & --81.58
  &  \\  
  incl. further $\delta_1,\delta_2,\delta_{12}$ terms:
  & 
  & \cite{SuPhD58,Ar57}
  & 9\ 849.58
  & \\
  \cline{1-4}\\[-0.3cm]
  -- incl. B$\cdot$C$\pupu$
  & $4(\frac{\pi}{2}+1) \delta_{12}$ 
  & (Su5.64)
  & 424.95$^\text{c}$
  & \\
  -- incl. T$\times$C$\pumi$   
  & $-4(1+\ln 2) \delta_{12}$ 
  & (Su5.21a)
  & --279.88
  & \\
  -- incl. T$\cdot$C$\mimi$   
  & $-4(\frac{\pi}{2}-1-\frac{1}{2}\ln 2) \delta_{12}$ 
  & (Su5.21b)
  & --37.06
  & \\
  \hline \\[-0.25cm]
  \textbf{Vacuum polarization} & $-\frac{4}{15}[Z (\delta_1+\delta_2)-\delta_{12}]$ & (Su7.20)
  &
  & \textbf{--176.59}
  \\   
  \hline\hline\\[-0.35cm]
  \multicolumn{2}{l}{%
  $\alpha^3\epsi_3^{\nulla}
  =
  \alpha^3
  \left\lbrace%
    \frac{4}{3}[-2\ln\alpha-\ln k_0+\frac{19}{30}]Z(\delta_1+\delta_2)
    +
    [\frac{14}{3}\ln\alpha+\frac{164}{15}]\delta_{12}-\frac{14}{3} Q
  \right\rbrace$} 
  & \cite{Ar57}
  &
  & \textbf{--7212.41} \\  
\hline\hline %
\end{tabular}}
\begin{flushleft}
$^\text{a}$ %
  References (equation numbers, labels) to Araki (1957) \cite{Ar57} and Sucher (1958) \cite{SuPhD58}. \\
$^\text{b}$ %
  We used $\alpha=1/137.035999084$, 
  $\delta_{12}=0.106\ 345\ 371\ 2$, 
  $Z(\delta_{1}+ \delta_{2})=1.810\ 429\ 318$ 
  (with $\delta_i=\langle\delta(\br_i)\rangle$)
  for the He 1S state \cite{Dr06}, $Q=0.0787238$ \cite{Dr88nucl}. \\
$^\text{c}$ %
  Automatically included in the no-pair DC(B) energy with an $\alpha$ dependence in agreement with the $\delta\epsi_3^\nulla$ nrQED correction \cite{JeFeMa22,FeJeMa22b}. Regarding $\text{T}\cdot\text{T}\pupu$, only the instantaneous part of the correction is included in the variational no-pair energy (see also footnote $^\text{c}$ to Table~\ref{tab:alphascaling}). 
\end{flushleft}  
\end{table}

First of all, we note that the (non-crossing) pair correction for the instantaneous interaction,
\begin{align}
  V_\delta
  =
  L_{++} V_\inst (1-L_{++})
  -
  L_{--} V_\inst
\end{align}
renders the Hamiltonian non-hermitian, though it remains energy-independent. Either direct solution of the instantaneous with-pair wave equation or a perturbative correction for this term to the no-pair energy are possible, detailed mathematical properties, and numerical results will be presented in Ref.~\citenum{JeMa24}.

Radiative, retardation, crossed-photon contributions, and other higher-order Feynman diagrams are carried by the non-linearity in the Salpeter--Sucher equation, Eq.~\eqref{eq:eBS2},
\begin{align}
  \Vcal_\epsilon(E)
  =
  (E-h_1-h_2)
  \int_{-\infty}^{+\infty} \frac{\dd\epsi}{-2\pi\iim}  
    s_1 s_2 \Kcal_\Delta
    \left[%
      1 - s_1 s_2 \Kcal_\Delta
    \right]^{-1}
    s_1 s_2
    V_\inst \; .
  \label{eq:Vcal1}
\end{align}
Prospects for explicit evaluation (resummation) of this term and an iterative solution of the full equation were recently contemplated in Ref.~\citenum{MaMa23}. 
Alternatively, one could attempt to solve iteratively Eq.~\eqref{eq:BS} or \eqref{eq:BSPsi}, by using the high-precision numerical solution of Eq.~\eqref{eq:Hnopair}.

In the present work, a perturbative correction to the no-pair energy is considered, which may be most straightforward at the lowest orders of perturbation theory. Using a `zeroth-order' approximation of $\Vcal$,  
\begin{align}
  \Vcale^{(0)} (E)
  =
  \qty(E-h_1-h_2)
  \int_{-\infty}^{+\infty}
  \frac{\dd{\epsi}}{-2\pi\iim}
      s_1 s_2 \Kcal_\Delta
      s_1 s_2 V_\inst \; ,
\end{align}
and then, the first-order energy correction can be obtained as \cite{MaFeJeMa23,MaMa23}
\begin{align}
  \Delta E^{(1,0)}_{\Kcal_\Delta}
  &\approx
  \langle 
    \Phi_\nopair |
    \int_{-\infty}^{+\infty}
    \frac{\dd{\epsi}}{-2\pi\iim}
      (s_{1+} + s_{2+}) \Kcal_\Delta
      (s_{1+} + s_{2+})
    \Phi_\nopair
  \rangle \; .
  \label{eq:pt10}
\end{align}
The most important diagrams to be included as $\Kcal_\Delta$ are highlighted in Figs.~\ref{fig:kernels}(d)--(m). Their importance to the helium atom ground-state energy can be estimated from Table~\ref{tab:nrQEDcorr}. 
For numerical, technical, and conceptual reasons, it will be the most important to start with the single-photon transverse or retardation correction, Figs.~\ref{fig:kernels}(g)--(h), the `crossed' Coulomb correction, `C$\times$C' Fig.~\ref{fig:kernels}(d), as well as the self-energy, the vertex, and the self-energy plus Coulomb ladder corrections, Figs.~\ref{fig:kernels}(m),~(k) and (i). Two of us have recently elaborated formal relations and carried out preparatory calculations with particular emphasis on the self-energy-related terms \cite{MaMa23}.

In the last paragraphs, we highlight some characteristics of the planned algorithmic and computational work. An important feature of the formalism is the presence of inherently one-particle quantities, \emph{e.g.,} the $s_1$ and $s_2$ propagators,  Eq.~\eqref{eq:pt10}. In special cases, it is possible to re-group the terms and factors 
during the course of the evaluation of the perturbative corrections, and genuine two-particle operators, \emph{e.g.,} $h_1+h_2$ reappear in the final expressions. However, this rearrangement is not generally possible. 

For a complete evaluation of the energy correction, including all relevant processes (Fig.~\ref{fig:kernels}), we must be able to tackle inherently one-particle operators. Is this a problem? Well, we can compute high-precision correlated (relativistic) energies using \emph{explicitly correlated} basis functions (Tables~\ref{tab:He1S_DCB}--\ref{tab:alphascaling}), whereas resorting to an orbital-based, \emph{i.e.,} \emph{single-particle} basis representation would result in several orders of magnitude loss of precision (back to m$\Eh$s instead of p$\Eh$s). This massive loss of precision was observed already during attempts to combine orbital-based projectors with explicitly correlated no-pair energy computations \cite{JeFeMa22}. 

Instead, it appears to be a better strategy to stay within the two-electron space spanned by the optimized two-electron functions (ECGs), and construct inherently one-particle operators over this two-particle space. In this case, a faithful representation can only be obtained if we consider the \emph{entire} Hilbert space, \emph{i.e.,} beyond the physically relevant anti-symmetrized subspace,  including also the mathematically necessary permutationally symmetric subspace of the Hilbert space.

\begin{figure}
\includegraphics[scale=1.]{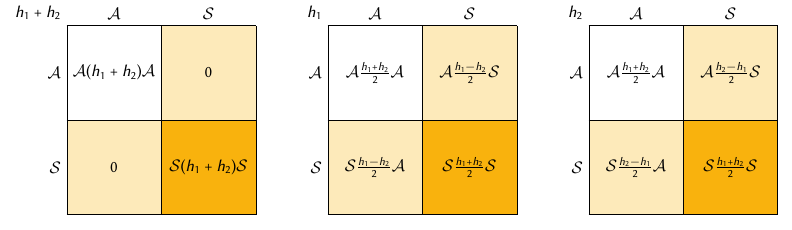}
\caption{%
  Representation of permutationally symmetric and non-symmetric operators over the entire (S, symmetric and A, anti-symmetric)  Hilbert space, for the example of the $h_1+h_2$ and $h_1$, $h_2$ Hamiltonians, respectively.
  \label{fig:permsym}
}
\end{figure}

Figure~\ref{fig:permsym} visualizes this idea for the example of the two-particle $h_1+h_2$,  the one-particle $h_1$, and the one-particle $h_2$ Hamiltonians. 
This procedure has already been successfully tested for the construction of the positive-energy projector and its application for the computation of no-pair DC(B) energies with an explicitly correlated basis set \cite{HoJeMa24}. 

Similar computational strategies will be developed for the matrix representation of inherently one-particle operators, \emph{e.g.,} one-particle propagators, occurring in QED corrections, \emph{e.g.,} Eqs.~\eqref{eq:Vcal1}--\eqref{eq:pt10}, in order to be able to benefit from the high numerical precision of the interaction energy ensured by an explicitly correlated basis representation.

\section{Summary, conclusion, and outlook}
We have reviewed applications of the equal-time Bethe--Salpeter equation for high-precision computations of two-electron atomic (and molecular) systems with potential future relevance for precision physics and spectroscopy. 

The equal-time Bethe--Salpeter (or Salpeter--Sucher) equation is an eigenvalue(-like) equation, in principle, equivalent to a full QED description of the bound-state system. Retardation, crossed-photon, self-energy or vacuum polarization `processes' can be accounted for through dressed propagators and vertices and appropriate interaction kernels. In the Salpeter--Sucher formalism, these quantities can be collected in an `effective' interaction kernel, and these contributions introduce a non-linear energy dependence into the wave equation. In QED, this non-linearity is estimated to be small, and hence, we first solve the linear, dominant part of the wave equation, which contains the no-pair Dirac--Coulomb(--Breit) Hamiltonian. The helium atom ground and first singlet excited state energies are demonstrated to be converged with a (sub-)n$\Eh$ precision in an explicitly correlated variational framework. This correlated, relativistic energy is orders of magnitude more accurate than the non-relativistic energy already for low nuclear charge values (\emph{e.g.,} $Z=2$), and hence, it provides a good starting point for the development of a \emph{relativistic} QED (rQED) perturbation theory. Furthermore, this rQED approach appears to have a mathematical structure simpler than that of nrQED, and carries only the `inherent' QED divergences corresponding to mass and charge renormalization, which will give us the opportunity to directly study these `strange' QED features and their implementation in a (high-precision) numerical procedure.

We have outlined initial directions for a perturbative account of the non-linear effective QED `potential' energy term and pointed out the ubiquitous presence of inherent one-particle quantities, for which we plan to construct a faithful representation by working over the entire Hilbert space (beyond its permutationally anti-symmetric subspace) spanned by an explicitly correlated basis representation. 

Extensions beyond two electrons (spin-1/2 fermions) appear to be possible 
along the lines outlined by Sucher \cite{Su80,Su83,Su84} and Broyles \cite{Br87}. We plan to elaborate on theoretical details and numerical procedures in future work.

\vspace{0.5cm}
\section*{Acknowledgements}
\noindent 
We thank Dávid Ferenc, Péter Jeszenszki, Péter Hollósy, and Eszter Saly for discussions and joint work over the past years. 
Financial support of the European Research Council through a Starting Grant (No.~851421) is gratefully acknowledged.

\vspace{0.5cm}
\section*{Supporting Information}
\noindent 
The ECG basis set parameters for the different basis set sizes 
and the $\alpha$ dependence of the DC and DCB energies used to prepare the tables are provided as Supplementary Information.

\vspace{0.5cm}
%

\end{document}